\pdfoutput=1
\documentclass[11pt,a4paper]{article}
\usepackage{jcappub}

\title{Reconstructing the inflationary $f(R)$ from observations}

\author[a,b]{Massimiliano Rinaldi,}
\author[a,b]{Guido Cognola,}
\author[a,b]{Luciano Vanzo,}
\author[a,b]{Sergio Zerbini.}

\affiliation[a]{Dipartimento di Fisica, Universit\`a di Trento\\
                                     via Sommarive 14, 38123 Trento, Italy.}
\affiliation[b]{TIFPA (INFN),\\ via Sommarive 14, 38123 Trento, Italy.}

\emailAdd{massimiliano.rinaldi@unitn.it}
\emailAdd{guido.cognola@unitn.it}
\emailAdd{luciano.vanzo@unitn.it}
\emailAdd{zerbini@science.unitn.it}

\newcommand{\beq}{\begin{equation}}
\newcommand{\eeq}{\end{equation}}

\newcommand{\be}{\begin{equation}}
\newcommand{\ee}{\end{equation}}

\newcommand{\bea}{\begin{eqnarray}}
\newcommand{\eea}{\end{eqnarray}}
\newcommand{\non}{\nonumber}

\abstract{
\noindent The BICEP2 collaboration has recently released data showing that the  scalar-to-tensor ratio $r$ is much larger than expected. The immediate consequence, in the context of $f(R)$ gravity,  is that the Starobinsky model of inflation is ruled out since it predicts a value of $r$ much smaller than what is observed. Of course, the BICEP2 data need verification, especially from Planck with which  there is some tension, therefore any conclusion seems premature. However, it is interesting to ask what would be the functional form of $f(R)$ in the case when the value of $r$ is different from the one predicted by the Starobinsky model. In this paper, we show how to determine the form of $f(R)$, once the slow-roll parameters are known with some accuracy. The striking result is that, for given values of the scalar spectral index $n_{S}$ and $r$, the effective Lagrangian has the form $f(R)=R^{\zeta}$, where $\zeta=2-\varepsilon$ and $|\varepsilon|\ll 1$. Therefore, it appears that the inflationary phase of the Universe is best described by a $R^{2}$ theory, with a small deviation that, as we show, can be obtained by quantum corrections.
}

 \arxivnumber{1406.1096}

\begin{document}

 \maketitle
 \flushbottom

%%%%%%%%%%%% INTRO  SECTION %%%%%%%%%%%%%%%%%%%%%
\section{Introduction}

\noindent In the light of several observational results obtained over the past decades, from COBE and WMAP to Planck, together with the recent BICEP2 observations \cite{bicep2}, the importance of single field inflationary models has been constantly sharpened, to the point that now it has taken the form of a scientific paradigm. These models are mathematically equivalent to modified gravity  with Lagrangians ${\cal L}=\sqrt{-\det(g)}f(R)$, provided one chooses as the relevant dynamical variable the metric field \footnote{The Palatini version of the theory does not fit very well with observational data.} 
(see \cite{review,defelice,Sotiriou:2008rp,fara,capo} for extensive reviews on the subject). 

Notwithstanding the lack of a general equivalence theorem, at least in the classical domain, they must also be considered  ``physically equivalent'', since the numerical values of the best tested observables is the same in both formulations, away from singularities \cite{Faraoni:1998qx,Deruelle:2010ht}. By contrast, the limited knowledge on the quantum aspects of these models falls short to a full equivalence, although the quantum corrections discussed in Sec.\ 3 look like the effective Coleman-Weinberg potential of scalar-tensor theories. As an example of the effects of field redefinitions, we recall that in flat space quantum field theory these lead to the same $S$-matrix. In any case, we believe that there are no reasons to suspect that the quantized linearized theory in either formulation are really different (for example, both have the same physical degrees of freedom). 

Virtually all models of modified gravity are considered in the form $f(R)=\frac{1}{2} M_p^{2}R\,+$ ``something else''. \footnote{Here, $M_{p}^{2}=8\pi G$ is the reduced Planck mass.} The most popular is the Starobinsky model, where the additional term reads $R^{2}/(6M^{2})$ and $M\simeq 10^{13}$ GeV  \cite{Starobinsky:1979ty,Starobinsky:1980te,defelice}.   In this paper we take the more radical step of omitting the linear Einstein-Hilbert term and keeping only the remaining stuff. The fundamental reason is that the linear term tends to suppress precisely the tensor-to-scalar ratio that we want to save from becoming too small in order to be consistent with the recent findings of BICEP2 \cite{bicep2}. Although there is a lot of discussion about the reliability of this experiment, our findings are, at the end of the day, independent from BICEP2. In fact, as we will show below, all we need is just a non-vanishing value of $r$. By assuming that the scalar spectral index $n_{S}$ and $r$ are given, we show that, in the slow-roll regime, there is  a unique and simple expression  for the dominant term of the effective Lagrangian, which reads $f(R)=R^{\zeta}$ where $\zeta=2-\varepsilon$, with $|\varepsilon|$ much smaller than one and decreasing for decreasing $r$. We also demonstrate that $\zeta$ is constant over a wide range of e-foldings that encompass the epoch of horizon exit of the relevant scales. We finally look for an explanation of the form of this term in the quantum corrections that are to be expected, according to the modern viewpoint of effective quantum field theory.  

The full structure of the effective Lagrangian is not in the class of attainable goals so that, in brief, our main idea is to restrict considerations to the observable number of e- foldings during inflation and start with a single term $R^2$, on the ground that, at the scale of inflation, conventional wisdom in quantum field theory would suggest a scale invariant world. Strictly speaking, the same requirement is also met by the quadratic Weyl action \footnote{Inflation with a Weyl term has been studied in \cite{Deruelle:2010kf}.}, and in fact the attractiveness of conformally invariant actions as fundamental actions for quantum gravity has been repeatedly stressed in the past \cite{Boulware:1983td}. However, as we mentioned above, single field inflation is the favorite model and this leads uniquely to $f(R)$ theories. We would like to note that a similar approach was recently proposed in \cite{strumia}. This model, dubbed ``Agravity'', is much more ambitious than ours as it requires that the fundamental theory of nature does not contain any physical scale.

The Einstein-Hilbert linear term hopefully emerges later in the cosmic evolution, as the result of a scale invariance symmetry breaking induced by matter quantum corrections \cite{Boulware:1983td,Adler:1982ri}. Eventually, this is the necessary mechanism to stop inflation and ignite reheating. This scenario is different from the model of induced gravity proposed long ago by Sakharov \cite{Sakharov:1967pk,Zeldovich:1967gd} in that the induced gravitational constant is renormalized and computable, so that no tree level linear term is needed. In practice, this means that $R^2$ is treated as the quantum side and $R$ as the classical emergent side of one and the same gravitational field. This is not an entirely heretical idea, given the notorious difficulties in formulating a quantum theory of the Einstein action, with the notable exception of loop quantum gravity.  

In the next section we discuss how all these considerations reduce, in our scheme, to a single differential equation for a function $\alpha(t)$,  given in terms of the slow-roll parameters, that fixes the exponent $\zeta$. The equation is numerically solved and the solutions are discussed to find the effective Lagrangian. In this context, we also offer a brief comparison with other models. In Sec.~3 we propose an explanation of our results in terms of a quantum corrected de Sitter solution of $R^2$ gravity. Finally, we conclude with some considerations in Sec. 4.

\section{Slow-roll parameters in Einstein and Jordan frames}

\noindent Let us consider the general $f(R)$ theory with Jordan-frame action
\bea\label{jaction}
S_{J}={M_{p}^{2}\over 2}\int d^{4}x\sqrt{-\det (g)}f(R)
\eea
in the vacuum and  the cosmological J-frame metric
\bea\label{Jmetric}
ds^{2}=-dt^{2}+a^{2}d\vec x^{2},
\eea 
where $a(t)$ is the scale factor. The equations of motion are
\bea\label{eom1}
3XH^{2}&=&{1\over 2}(XR-f)-3H\dot X,\\\label{eom2}
\ddot X&=&-2X\dot H+H\dot X,
\eea
where the dot refers to a derivative with respect to the (J-frame) cosmic time $t$, $H=a^{-1}\dot a$ is the Hubble function, and
\bea\label{definitions}
X\equiv {df(R)\over dR}, \qquad R=6(2H^{2}+\dot H).
\eea
Our scope is to write the E-frame slow-roll parameters in terms of J-frames quantities only and for a generic $f(R)$. One possible procedure is to write first the theory as an equivalent scalar-tensor theory in Einstein frame and then compute the slow-roll parameters as usual. The problem is, however, that it is not alway possible to follow this strategy, for example when $f(R)$ contains terms like $\ln ( R)$, as it happens when loop corrections are taken in account.

However, no matter the form of $f(R)$, we know that the conformal transformation between Jordan and Einstein frames has the form $\tilde g_{\mu\nu}=Xg_{\mu\nu}$, where, from now on, tilded quantities refer to the Einstein frame. It has been proven that the spectrum of cosmological perturbations is invariant under conformal transformation (see e.g.  \cite{defelice}). As a consequence, also the observed spectra are invariant and so are the spectral indices. We can exploit this feature to write an observed spectral index as a function of E-frame or J-frame quantities. In particular, we can avoid to write the $f(R)$ theory as an explicit scalar-tensor theory and fulfill the aim of expressing the spectral indices as functions of $H$ and $X$ only.

In E-frame the metric reads
\bea
d\tilde s^{2}=-d\tilde t^{2}+\tilde a^{2}d\vec {\tilde x}^{2},
\eea
from which it follows that
\bea
\tilde a =\sqrt{X}a, \quad {dt\over d\tilde t}={1\over \sqrt{X}},
\eea
and
\bea
\tilde H={1\over \tilde a}{d\tilde a\over d\tilde t}={dt\over d\tilde t}\,{d\over dt}\left[ \ln\left(a\sqrt{X}\right)\right]={1\over \sqrt{X}}\left({\dot X\over 2X}+H\right).
\eea
By differentiating again with respect to $\tilde t$ and by using the  equation of motion \eqref{eom2} we find
\bea
{d\tilde H\over d\tilde t}=-{3\dot X^{2}\over 4X^{3}},
\eea
and
\bea
{d^{2}\tilde H\over d\tilde t^{2}}={1\over \sqrt{X}}\left({9\dot X^{3}\over 4X^{4}} +{3\dot X\dot H\over X^{2}}-{3H\dot X^{2}\over 2X^{3}}  \right).
\eea
By combining these expressions, we find that the first two Hubble flow functions, defined by Eqs.\ \eqref{Hubbleflow} in the Appendix, can be written as
\bea\label{epsilon1}
\tilde \epsilon_1 ={3\dot X^{2}\over (\dot X+2XH)^{2}},
\eea
and
\bea\label{epsilon2}
\tilde \epsilon_{2}=-{8X(H\dot X^{2}+X\dot X\dot H+2X^{2}H\dot H-X\dot X H^{2})\over \dot X(\dot X+2HX)^{2}}.
\eea
The function $\tilde\epsilon_{3}$ is quite complicated and we do not report it. It will be shown in a simplified form below.
The expression \eqref{epsilon1} can always be inverted and written as
\bea\label{dotalpha}
\dot X=\alpha HX, \quad \alpha={2(\tilde\epsilon_{1}\pm\sqrt{3\tilde\epsilon_{1}})\over 3-\tilde\epsilon_{1}}.
\eea
By definition, accelerated expansion in Einstein frame occurs as long as  $|\tilde\epsilon_{1}|<1$, namely when $1-\sqrt{3}<\alpha<1+\sqrt{3}$. During inflation, it is known that $\tilde \epsilon_{1}$ is not constant thus $\alpha$ is a function of time. By keeping this in account, from the equation of motion \eqref{eom2} we obtain the (exact) relation
\bea\label{dotH}
\dot \alpha H+(\alpha+2)\dot H=\alpha(1-\alpha)H^{2}.
\eea
In terms of the number of e-folding $N$ (defined as $dN=Hdt$), the expression above becomes
\bea\label{Hprime}
(\alpha+2){H'\over H}+\alpha(\alpha-1)+\alpha' =0,
\eea
while \eqref{dotalpha} becomes simply $X'=\alpha X$ (from now on, a prime will indicate a derivative with respect to $N$). Using these equations, it easy to express the first three Hubble flow functions in terms of $\alpha$ and its N-derivatives only as
\bea
\tilde\epsilon_{1}&=&{3\alpha^{2}\over (\alpha+2)^{2}},\\
\tilde\epsilon_{2}&=&{8\alpha'\over \alpha(\alpha+2)^{2}},\\
\tilde\epsilon_{3}&=&-{2(3\alpha+2)\alpha'\over \alpha(\alpha+2)^{2}}+{2\alpha''\over (\alpha+2)\alpha'}.
\eea

By using the equations of motion, the definitions \eqref{definitions}, and the equation \eqref{Hprime}, we finally find that, on the cosmological background \eqref{Jmetric}, the $f(R)$ theory is effectively equivalent to
\bea\label{Rzeta}
f(R)=f_{0}R^{\zeta}\, ,\qquad \zeta={4-\alpha-\alpha'\over 2(1-\alpha)-\alpha'}\,,
\eea
where $f_{0}$ is an arbitrary constant. Of course, this cannot be taken as a fundamental action generating field equations, except for when $\zeta$ is constant. However, it can be taken as a numerical value of the Lagrangian density characterizing the scale of inflation. We remind that $\zeta=2$ corresponds to a scale invariant theory while $\zeta=1$ leads to the usual Einstein-Hilbert term.

The expressions \eqref{Rzeta} are exact and we wish to determine $\alpha$ during inflation by means of observational data. This is possible, of course,  because the Hubble flow functions are related to the slow-roll parameters in the slow-roll approximation. Since we can determine the slow-roll parameters, through the measure of $\tilde n_{S}$ and $\tilde r$ in Einstein frame, we can also determine the function $\zeta(N)$. 

Suppose first that $\alpha$ is constant. In this case, we find 
\bea\label{eps1}
\tilde\epsilon_{1}={3\alpha^{2}\over (\alpha+2)^{2}},\qquad \tilde\epsilon_{2}=0.
\eea
In particular, the second equality implies that $2\tilde\epsilon_{V}=\tilde\eta_{V}$ and  eq.\ \eqref{ns}  that
\bea\label{nsalpha}
\tilde n_{s}={4+4\alpha-5\alpha^{2}\over (\alpha+2)^{2}}.
\eea
By using the Planck value $\tilde n_{s}=0.9603\pm 0.0073$ \cite{planck}, we  can fix $\alpha$ and, in turn, the tensor-to-scalar ratio by means of eq.\ \eqref{ratio}, which yields $0.259<\tilde r<0.376$. We notice that the lowest value for $\tilde r$ is marginally compatible with the recent findings of BICEP2 \cite{bicep2}. Although probably too large, this result is useful as it helps us to check that the solutions to the equations of motion correctly reproduce inflation. In fact, if $\alpha$ is constant  we can integrate the equations of motion exactly. The Hubble parameter turns out to be
\bea
H(t)={\alpha+2\over \alpha(\alpha-1)(t-t_{0})}\Longrightarrow a(t)\sim (t-t_{0})^{{\alpha+2\over \alpha(\alpha-1)}}
\eea
for arbitrary $t_{0}$. We checked that the two possible central values of $\alpha$ obtained by solving eq.\ \eqref{nsalpha}, correspond to one expanding and one contracting Universe, according to
\bea
\alpha&=&-0.150 \Longrightarrow a(t)\sim t^{10.7}\,,\\
\alpha&=&+0.177 \Longrightarrow a(t)\sim t^{-14.95}\,.
\eea
This information is important to remove the degeneracy that will shop up below, when we will take a time-dependent $\alpha$.

As a further check of our findings, we transform the $f(R)$ expression found above into the corresponding tensor-scalar theory. The associate scalar potential has the general form
\bea
V(\tilde\phi)={XR-f(R)\over 2\kappa^{2}X^{2}},
\eea
while the (Einstein frame) scalar field $\tilde\phi$ is identified through the relation
\bea
X=\exp\left(\kappa\sqrt{2\over 3}\tilde\phi\right).
\eea
By using the results above, we find that the potential can be written as
\bea
V=V_{0}\exp \left[-\kappa \sqrt{2\over p}\,\tilde\phi\right]
\eea
for some constant $V_{0}$ and for (see eq.\ \eqref{eps1})
\bea
p={(\alpha+2)^{2}\over 3\alpha^{2}}={1\over \tilde \epsilon_{V}}.
\eea
The potential is the one proposed by Lucchin-Matarrese in \cite{lucchin} and the fact that $p$ is the inverse of the slow-roll parameter confirms that the solution is of  power-law type also in Einstein frame and has the form $\tilde a\sim \tilde t^{p}$.

\begin{figure}[tbp]
\centering
\includegraphics[width=7cm]{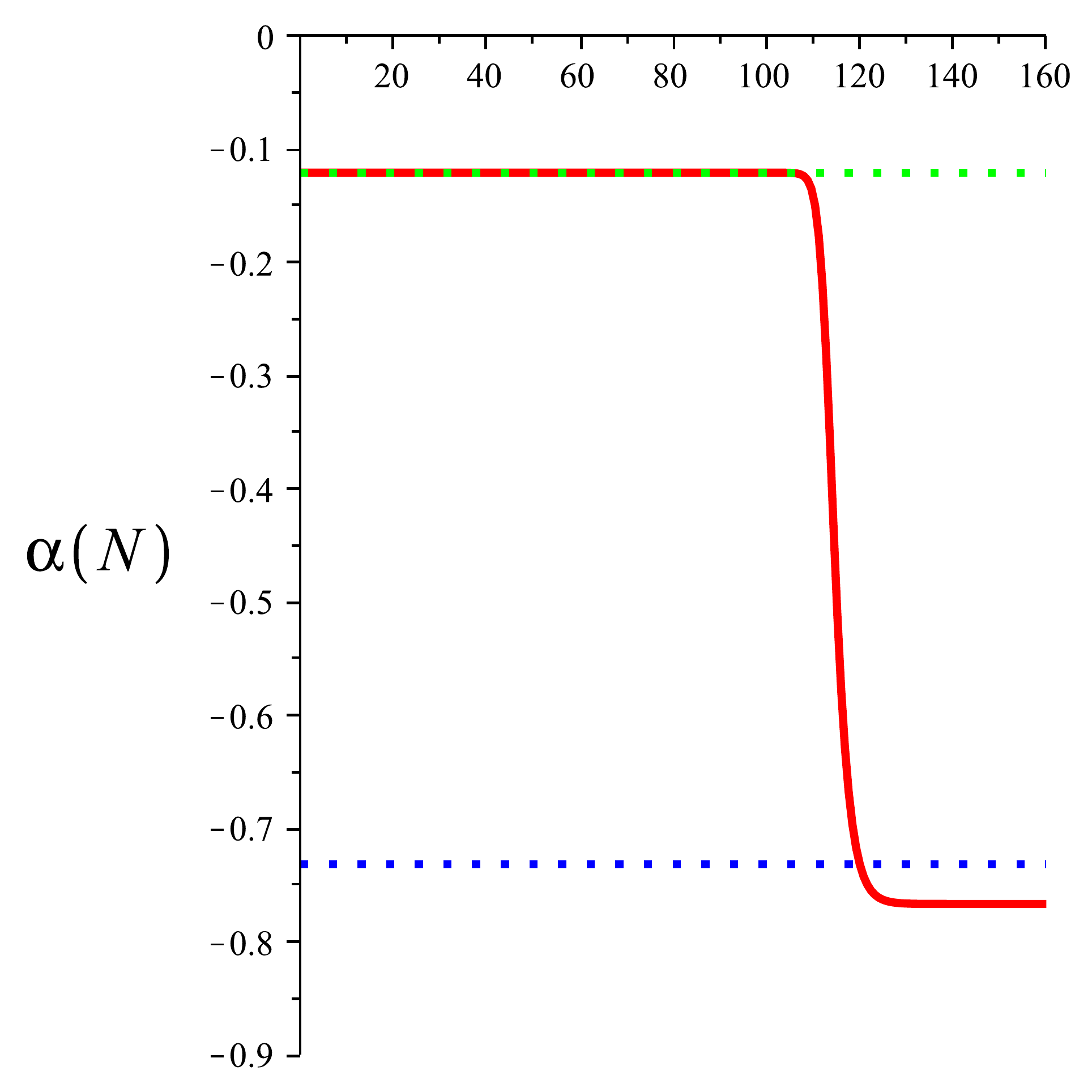}
\caption{\label{plot1} Plot of $\alpha(N)$ (red curve) with the initial condition $\alpha(120)=1-\sqrt{3}$. The green dotted curve is the asymptotic value $\alpha\simeq -0.1211$ and the blue one corresponds to $\alpha=1-\sqrt{3}$.}
\end{figure}

\begin{figure}[tbp]
\centering
\includegraphics[width=7cm]{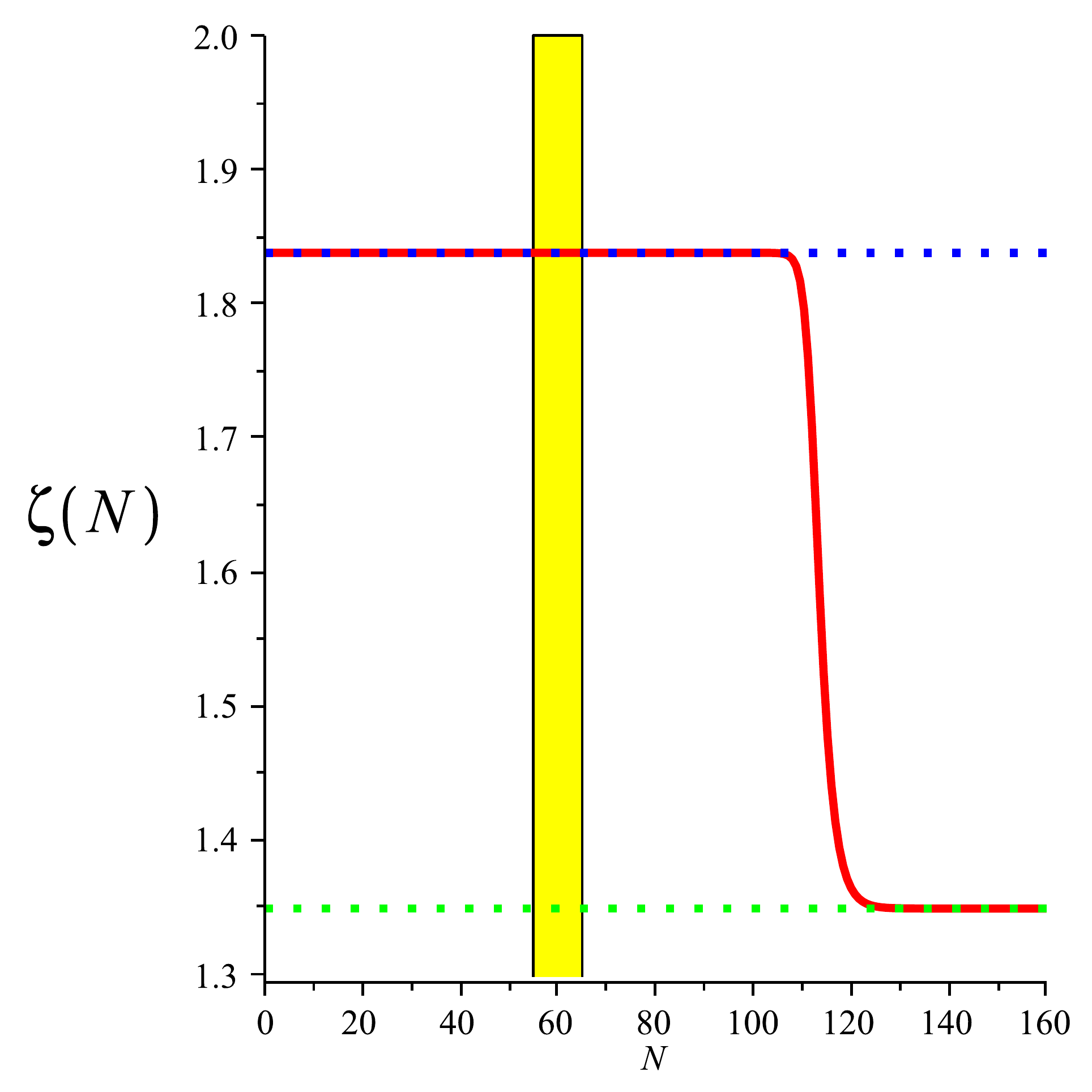}
\caption{\label{plot2} Plot of the exponent $\zeta(N)$ (red curve) of $R^{\zeta}$. The green and the blue lines are the asymptotic values $\zeta\simeq 1.3489$ and $\zeta\simeq 1.8380$ respectively. The yellow band corresponds to the region when relevant scales exit the Hubble horizon.}
\end{figure}

We now look at the more realistic case of a time-dependent $\alpha$ during inflation. With the help of Eqs.\  \eqref{ns} and \eqref{ratio}, we can write the scalar spectral index and the tensor-to-scalar ratio respectively as 
\bea\non\label{nshigher}
\tilde n_{S}-1&=&{16C(3\alpha+2)\alpha'^{2}\over \alpha^{2}(\alpha+2)^{4}}-{16(2+2\alpha+(3C+5)\alpha^{2})\alpha'\over \alpha(\alpha+2)^{4}}-{16C\alpha''\over \alpha(\alpha+2)^{3}}-{24\alpha^{2}(\alpha^{2}+\alpha+1)\over (\alpha+2)^{4}},\\
\eea
and
\bea\non
\tilde r&=&{
\frac {48{\alpha}^{2}}{ \left( \alpha+2 \right) ^{2}}}
-{\frac {32 \left( 10\,{\pi }^{2}-24\,{C}^{2}+36\,{C}^{2}\alpha-96
-3\,{\pi }^{2}\alpha \right) {\alpha'}^{2}}{ \left( \alpha+2 \right) 
^{6}}}+\\
&+&{\frac {192\alpha\, \left( 8\,C+8\,C\alpha-3\,{\pi }^{2}{
\alpha}^{2}+8\,C{\alpha}^{2}+30\,{\alpha}^{2} \right) {\alpha'}}{
 \left( \alpha+2 \right) ^{6}}}+{\frac {32\alpha\, \left( 12\,{C}^{2
}-{\pi }^{2} \right) {\alpha''}}{ \left( \alpha+2 \right) ^{5}}}.\label{rhigher}
\eea
If we consider $\tilde r$ and $\tilde n_{S}$ nearly constant during inflation (and determined by experiments) we can numerically solve the system of differential equations \eqref{nshigher} and \eqref{rhigher}.\footnote{Note that Eq.\ \eqref{nshigher} does not coincide with Eq.\ \eqref{nsalpha} when $\alpha=$ const. The reason is that the higher order expression for $n_{S}$ includes the term $\tilde\epsilon_{1}^{2}$ that does not, however, contain derivatives of $\alpha$ but is still considered being of higher order than $\tilde\epsilon_{1}$ and $\tilde\epsilon_{2}$, see eq.\ \eqref{ns}.} The combination of the two equations above yields a first order differential equation, which is quadratic in $\alpha'$ therefore there are two solutions. Thanks to the results obtained above in the approximation of constant $\alpha$, we know, however, which one corresponds to an expanding Universe, that is the negative one. The constant of integration is fixed assuming that  inflation ends at a given $N_{\rm end}$, such that $\alpha(N_{\rm end})=1-\sqrt{3}$, see Eq.\ \eqref{dotalpha}. The results are plotted in Figs.\ \eqref{plot1} and \eqref{plot2} in the case when we take the Planck central value $\tilde n_{S}=0.9603$ and the BICEP2 recent finding $\tilde r =0.2$. 

This is the first important result of this paper: the system of non-linear differential  equations $(\tilde n_{S}, \tilde r)$ yields a solution with an almost constant $\alpha$ (hence $\zeta$) except for a sharp transition (to a lower $\zeta$) towards the end of inflation. The result is accurate since the measured values of $\tilde n_{S}$ and $\tilde r$  are related to the spectra of fluctuations exiting the Hubble horizon about 60 e-folds before the end of inflation at $N=N_{\rm exit}$. The corresponding scales all exit the horizon in a range of about 10 e-folds centered around $N_{\rm exit}$  \cite{liddle}. In this interval, we see that $\alpha$ is extremely close to a constant. By changing the value of $\tilde n_{S}$ to keep in account Planck error bars modifies the asymptotic values of $\alpha$ of a negligible amount. 

\begin{figure}[tbp]
\centering
\includegraphics[width=8cm]{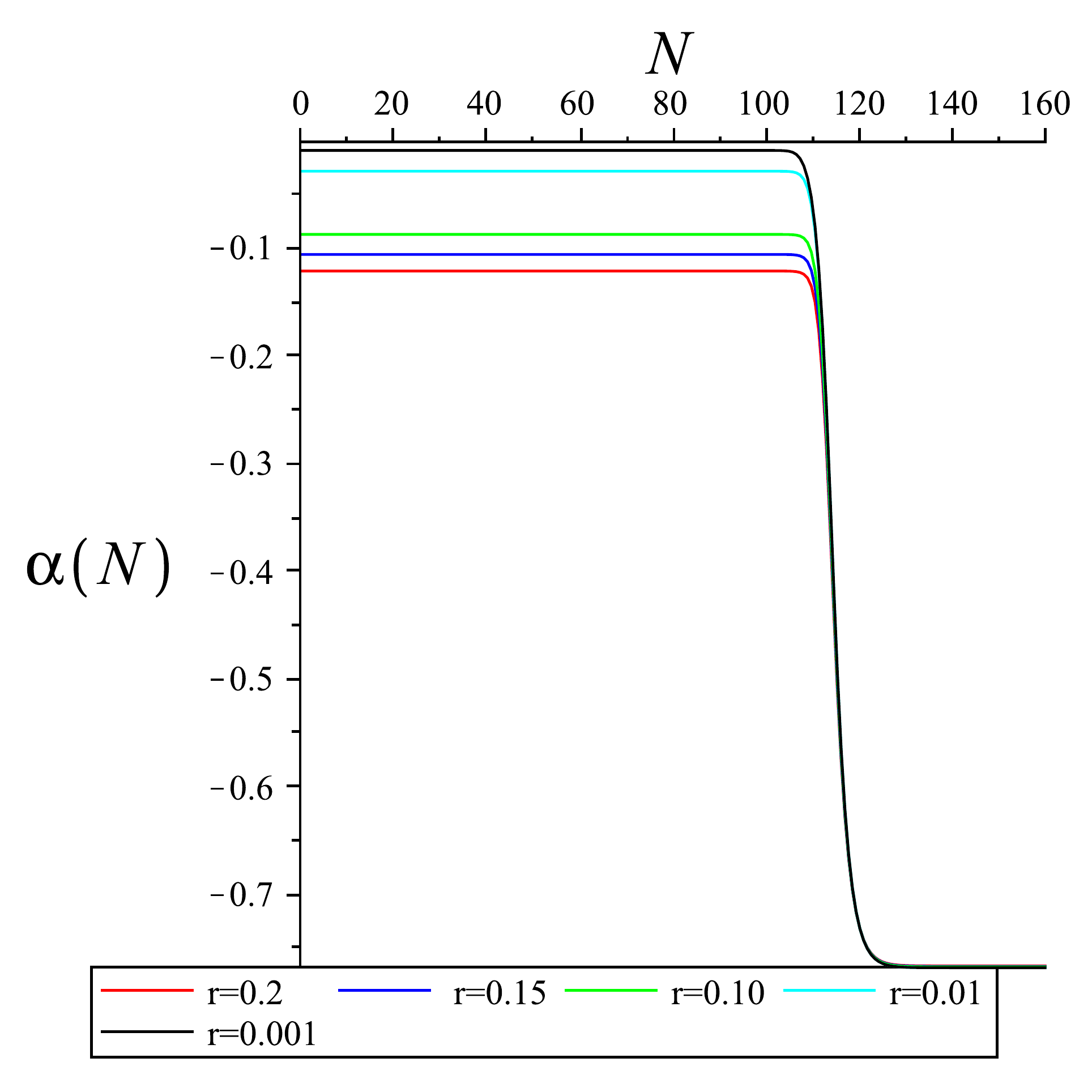}
\caption{\label{fig3} Plot of  $\alpha(N)$ for various values of $\tilde r$ and with the same initial condition as in Fig.\ \eqref{plot1}.}
\end{figure}

\begin{figure}[tbp]
\centering
\includegraphics[width=8cm]{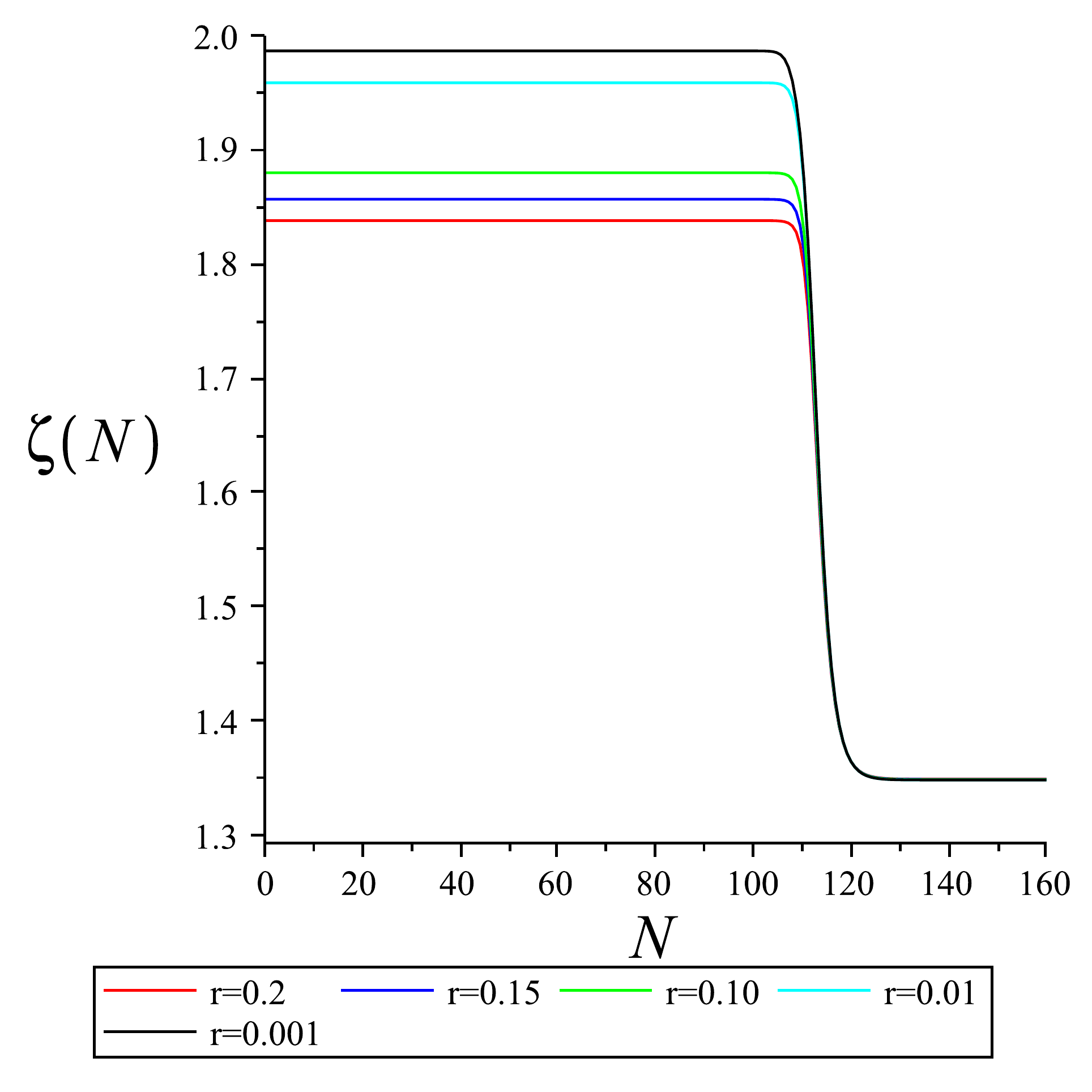}
\caption{\label{fig4} Plot of  $\zeta(N)$ for various values of $\tilde r$ and with the same initial condition as in Fig.\ \eqref{plot1}. It is evident that, as $\tilde r\rightarrow0^{+}$,  $\zeta\rightarrow 2^{-}$.}
\end{figure}

One might suspect that our result holds only if the BICEP2 measure of $\tilde r$ is correct. However, if we change the value of $\tilde r$ we find exaclty the same qualitative behaviour but with different asymptotic values for $\alpha$ and $\zeta$. In particular,  we see that if  $\tilde r$ decreases then $\alpha(N_{\rm exit})$ tends to $0^{-}$ (see Fig.\ \eqref{fig3}) while $\zeta$ grows towards $\zeta=2$ (see Fig.\ \eqref{fig4}). This is consistent with the fact that in de Sitter space (that would be one of the possible cosmological solutions of $R^{2}$ theory)  one has $\tilde \epsilon_{V}=0$ and, for single-filed inflation, $\tilde r=16\tilde\epsilon_{V}=0$. This is the second important result of this paper: the exponent $\zeta$ is uniquely fixed by the values of $\tilde n_{S}$ and $\tilde r$. We note that our model is phenomenologically equivalent to the Starobinsky one when the tensor-to-scalar ration has the value predicted by the latter, namely $\tilde r=0.00386$ \cite{planck}.

By the same token, we see that raising the value of $\tilde r$ tends to lower the exponent $\zeta$ monotonically. This observation helps to compare our results with the recent work \cite{Chakravarty:2014yda}, where a model $R+R^{\beta}$ is investigated yielding results comparable to ours. At $60$ e-folds before the end of inflation and with a  best fit value of $\beta\simeq1.818$, the authors find the value $\tilde r=0.264$, whereas our asymptotic value for $\tilde r=0.2$ is $\zeta=1.838$, in agreement with the observed monotonicity of $\zeta(N,\tilde r)$ versus $\tilde r$ for fixed $N$. It is to be noted that in \cite{Chakravarty:2014yda} the Einstein-frame potential is approximated to match exactly our model, because the Einstein-frame scalar field is assumed to take values much larger than the Planck mass.  The small numerical discrepancy is mostly due to the fact that, in our calculations, we used up the the third Hubble flow functions, which corresponds to the order $\tilde \epsilon_{V}^{2}$ (see the details in the appendix), while the authors of \cite{Chakravarty:2014yda} considered only the order $\tilde \epsilon_{V}$. The other notable result in \cite{Chakravarty:2014yda} is the relation of the model with SUGRA models, whereas our assumption is to trace its origin to the (necessarily present) quantum corrections affecting scale invariant gravity, in a viewpoint similar to that of Salvio and Strumia \cite{strumia}. 

We can also compare our findings with the well-known Higgs inflationary models \cite{Bezrukov:2007ep,Barvinsky:2008ia,DeSimone:2008ei,Barvinsky:2009fy,Barvinsky:2009ii}. In the former versions, the lower bound on the Higgs mass \footnote{This is $2\sigma$ away the value measured by LHC.}, necessary to the existence of inflation, together with the form of the effective potential, imply a value of $\tilde r$ much smaller than the one found by BICEP2. However, as recently observed in \cite{Hamada:2014iga,Bezrukov:2014bra}, near  the critical value of the Higgs mass the radiative corrections change the effective potential in such a way that $\tilde r$ is sensibly larger. Even if the Standard Model (SM), implemented with Einstein gravity, is non-renormalizable, this is a conceptually important observation as it saves the theory from being ruled if the experimental value of $\tilde r$ is higher than the one predicted by the ``bare'' Higgs inflationary model (or, equivalently, by the Starobinsky model). In the present case, we stress that our model is in principle independent from the SM parameters because the source of inflation is, in fact, just an extra degree of freedom encoded in the function $f(R)$. In principle, this model can be always transformed into an equivalent  single-field action. Then, the non-renormalizability of this theory is reflected in the non-renormalizability of the corresponding inflaton potential, which is required to be a general function since the very large values of the field during inflation are not suppressed by the Planck mass. On the other hand, the inflationary slow-roll parameters are computed from the essentially free theory which governs the small fluctuations around the background. Loop corrections can be computed in the ``in-in'' formalism and can be meaningful even for non renormalizable theories (see, e.g., Ref.~\cite{Weinberg:2005vy}).

\section{\texorpdfstring{$R^{2}$}{} and its one-loop quantum corrections} 

\noindent In the previous sections, we have shown how  the simple effective Lagrangian $R^\zeta$, with $\zeta\leq 2$, emerges from the observational data. In this section, we would like to justify our findings by means of one-loop quantum corrections.   The underlying idea is quite old since it was already contained in the seminal paper \cite{Starobinsky:1980te}, where quadratic terms in the curvature, justified by quantum effects, were added to the Einstein-Hilbert Lagrangian (for a review, see \cite{buch}).
One important recent finding is that the inclusion of suitable higher order contributions 
may realize not only the current accelerated expansion, but also  the inflationary epoch \cite{seba08}.
Within this context, the de Sitter (dS) space-time plays a fundamental role,
being able to provide an acceleration at different stages in the cosmological set up. Furthermore, modified gravity with quantum correction (trace anomaly), which may compatible with  BICEP2, was studied in \cite{bamba}. Further related investigations can be found in \cite{mazu,Costa}.

In  previous papers \cite{cogno,guido2,cogno9}, $f(R)$ gravity models at one-loop level in a de Sitter
background have been investigated.
A similar program for the case of  pure Einstein gravity was initiated in
Refs.~\cite{perry,duff80,frad} (see also \cite{ds1,vass92}). Furthermore, such approach
 also suggests  a possible way of understanding the cosmological constant
issue \cite{frad}. Hence, the study of one-loop generalized modified gravity  is a
natural step to be undertaken for the completion of such program,
with the aim to better understand the role and the origin of quadratic corrections in the curvature.
An alternative approach, which is in some sense alternative, has been proposed by Reuter and collaborators \cite{mart}, see also  the review papers 
\cite{martin} and \cite{martin1,percacci}, where quantum gravity effects in astrophysics and  cosmology are studied. The quantization of the Einstein-Hilbert theory plus quadratic curvature terms has been discussed in many papers, beginning with the detailed study in flat space presented in the seminal paper \cite{stelle}.
A preliminary discussion of a quadratic model based on one-loop on-shell results has been presented in \cite{cognola12}. Recently, a deformation of Starobinsky models has also been discussed also in \cite{cod}.

In our case, the starting point is complete different, since  the Einstein-Hilbert term will be not considered, and all degrees of freedom are the ones related to $R^2$. Let us first revisit the classical solutions associated with the Lagrangian $\sqrt{-\det(g)}R^{2}$ (see also \cite{per}). The condition for the existence of de Sitter solution, $2f(R)=RX(R)$, is automatically satisfied by the equations of motion \eqref{eom1} and \eqref{eom2}. Thus,  we have the de Sitter solution  
\beq
ds^2=-dt^2+\exp{\left(2t\over t_{0}\right)} d^2 \vec x\,,
\eeq
with arbitrary $t_0$ and nonvanishing constant curvature $R_0=\frac{12}{t_0^2}$. This solution will be our background solution for the calculation of quantum corrections.  On the other hand, $R=0$ is also  a  solution, which may yield Minkowski space but also  a radiation-dominated Universe with scale factor  $a(t)=a_0t^{1/2}$. Finally, there exist other solutions, discussed in \cite{per} and  
\cite{clift}, that satisfy the non-linear differential equation
\beq
2H\ddot H-\dot H^2+6H^2 \dot H=0\,,
\eeq
but that will not be  investigated here, because we are interested in the one-loop quantum 
corrections to the  classical de Sitter solution. In passing, we also observe that the Schwarzschild metric is a spherically symmetric static solution of $R^2$.

For the quantum corrections, we consider the classical Euclidean gravitational action
\beq
I_E[g]=-\int\:d^4x\,\sqrt{\det(g)}\,f(R)=-\frac{b}2\,\int\:d^4x\,\sqrt{\det(g)}\, R^2\,,
\label{action0}\eeq
where $b$ is a dimensionless arbitrary constant. At the classical level, its role is irrelevant, but, as we will see, it becomes crucial at the quantum level. We also note that the $R^2$ Lagrangian has the important property that the one-loop off-shell effective action coincides with the on-shell one, since  the off-shell factor  $2f-RX$  identically vanishes. Furthermore, for this model the ``scalaron'' mass is also vanishing, 
and  it is possible to show, by making use of the techniques presented in  \cite{cogno,guido2,cognola13},  that the $R^2$ action is one-loop renormalizable, the counterterm  being generated by the ``bare'' dimensionless parameter $b$. As a consequence, the renormalized $b(\mu)$ is running and, at one-loop level, it reads
\beq
b(\mu)=\gamma\,\ln\left(\frac{\mu^2}{\mu_0^2}\right)+b_0\,, \quad \gamma=-\frac{25}{1152\pi^2}\,.
\eeq
Similar quantum corrections have been investigated in \cite{Ben}.

With regard to this result, a clarification seems to be necessary. In fact, it is well known that, when working in an arbitrary background, the complete renormalizability requires also the $R_{\mu\nu}R^{\mu \nu}$ term, which, in our de Sitter background, is  proportional to $R^2$. However, what we have in mind here a particular case of the general scale-invariant action with $\mathcal{L}\sim\alpha R^{2}+\beta C_{\mu\nu\sigma\rho}C^{\mu\nu\sigma\rho}$. This is asymptotically free, it has zero energy for any asymptotically flat initial data when $\alpha,\beta\geq 0$ \cite{Boulware:1983td}, and it is probably renormalizable in the sense that it only requires scale-invariant counterterms.  Whether or not we can really take $\beta=0$ without destroying these properties is an open question that we hope to answer soon.

Returning to the main point, the discussion above leads to the one-loop effective Lagrangian  
\beq
L=-\frac{b_0}2\, R^2\left(1+\frac{\gamma}{b_0}\ln\frac{R^2}{\mu^2_0} \right)\,.
\eeq
If the free parameter $b_0$ is chosen such that $\frac{\gamma}{b_0}\ll1$, one finds that
\beq
L\sim -\frac{b_0}2\, R^2 \left(\frac{R^2}{\mu^2_0}\right)^{\frac{\gamma}{b_0}}\,,
\eeq
which provides a possible explanation of the phenomenological modified gravitational Lagrangian found in the previous section. From Eq.\ \eqref{Rzeta}, we see that we can write
\bea
\zeta=2-{\alpha'+3\alpha\over \alpha'+2\alpha-2},
\eea
and we have shown that observational data constrain the second term to be much smaller than unity. We then propose that this deviation from $\zeta=2$ is generated by the one-loop correction, namely that  
\bea
-{\alpha'+3\alpha\over \alpha'+2\alpha-2}=\frac{2\gamma}{b_0}.
\eea
In fact, $\gamma<0$ and and $b(\mu)$ is a decreasing running coupling constant, compatible  with the asymptotic freedom of quadratic gravity \cite{Fra,Tombo,avra,avra1}. For example, for $\tilde r=0.2$ we have $b_{0}=0.027$ while, for $\tilde r=0.001$, we find $b_{0}=0.106$. In general, the value of $b_{0}$ increases for decreasing $\tilde r$. This result implies the intriguing possibility of measuring almost directly quantum gravitational effects by means of cosmological observations.

%%%%%%%%%%%%%%%%%%%%%%%%%%%%%%%%%%

\section{Discussion}

%%%%%%%%%%%%%%%%%%%%%%%%%%%%%%%%%%

\noindent We have shown that the assumption of slow-roll evolution during inflation uniquely fixes the form of $f(R)$, yielding a non-integer power of $R$, namely $R^\zeta$, with $\zeta$ slightly smaller than $2$. This expression does not reduce to the Starobinsky model in any limit. We have also shown that a possible explanation for this result comes from one-loop quantum corrections to the classical Lagrangian $R^2$ that can account for the deviation from the scale-invariant case $R^{2}$.  Our results relies only upon  the value for the scalar perturbation index measured by Planck and on the hypothesis that the scalar-to-tensor ratio is nonvanishing (including the value found by BICEP2).

The picture that emerges is of an inflationary Universe governed by a scale-invariant extension of general relativity together with quantum corrections. In opposition to the Starobinsky model, our proposal does not have a built-in mechanism that can terminate inflation. We are confident, however, that inflation can be stopped in some other way and we hope to report soon new results in this direction.

%%%%%%%%%%%%%%%%%%%%%%%%%%%%%%%%%%%%
%%%%%%%%%%%%%%%%%%%%%%%%%%%%%%%%%%%%

\acknowledgments
We thank R.\ Percacci for valuable private communications, G.P.\ Vacca for comments and suggestions, and all the participants to the First Flag Meeting ``The Quantum and Gravity'' (Bologna, May 28-30, 2014) for stimulating discussions.

%%%%%%%%%%%%%%%%%%%%%%%%%%%%%%%%%%%%
%%%%%%%%%%%%%%%%%%%%%%%%%%%%%%%%%%%%

\appendix

\section{Hubble flow functions and slow-roll parameters}
\noindent We recall the definitions of Hubble flow functions \cite{ency}
\bea
\epsilon_{0}={H_{i}\over H},\quad \epsilon_{i+1}={\dot \epsilon_{i}\over H\epsilon_{i}},
\eea
where $H_{i}$ is the initial value of the Hubble function. In particular, we are interested in 
\bea\label{Hubbleflow}
\epsilon_{1}&=&-{\dot H\over H^{2}}=-{H'\over H}\,,\\\nonumber
\epsilon_{2}&=&{\ddot H\over H\dot H}+2\epsilon_{1}={H''\over H'}-{H'\over H}\,,\\\nonumber
\epsilon_{3}&=&{d\ln\epsilon_2(N)\over dN}\,,
\eea
where the prime stands for the derivative with respect to $N$, the e-fold number, determined by $dN=Hdt$. Note that these are \emph{exact} expressions.

In single-field inflation, with a generic potential $V(\phi)$, $\phi$ being the inflaton field, the relevant equations of motion are (here $8\pi G=1$) 
\bea
3H^2=\frac12\dot \phi^2+V, \quad \ddot \phi+3H\dot \phi+V_{\phi}=0, 
\eea
where $V_{\phi}={dV\over d\phi}$.
For suitable potentials, there exists a slow-roll regime such that  the kinetic term and the second time derivative of the inflaton can be neglected, so that
\bea
3H^2\simeq V\, , \quad 3H\dot \phi \simeq -V_{\phi}\,.
\eea
In this regime, we find that
\bea
\epsilon_{1}&\simeq& \frac12\left(V_{\phi}\over V\right)^{2},\\
\epsilon_{2}&\simeq& 2\left({V_{\phi}^{2}\over V}-{V_{\phi\phi}\over V} \right),\\
\epsilon_{2}\epsilon_{3}&\simeq&2\left[ {V_{\phi\phi\phi}V_{\phi}\over V^{2}}-3{V_{\phi\phi}\over V}\left(V_{\phi}\over V\right)^{2}+2\left(V_{\phi}\over V\right)^{4} \right].
\eea
If we define the slow-roll parameters (valid in the slow-roll regime only) as
\bea
\epsilon_{V}=\frac12\left(V_{\phi}\over V\right)^{2},\quad \eta_{V}={V_{\phi\phi}\over V},
\eea
we can, at the lowest order, write the relations
\bea
\epsilon_{1}\simeq \epsilon_{V},\quad \epsilon_{2}\simeq4\epsilon_{V}-2\eta_{V}.
\eea
We recall that all exponential potentials of the form $V\sim \exp(k\phi)$ yield $\eta_{V}=2\epsilon_{V}$ for all constants $k$.

For the calculations shown in the main text, we are interested in the expressions that link directly the slow-roll parameters to the Hubble flow functions. One can show that \cite{Martin:2006rs}
\bea\label{ns}
n_{S}&\simeq&1-2\epsilon_{1}-\epsilon_{2}-2\epsilon_{1}^{2}-(2C+3)\epsilon_{1}\epsilon_{2}-C\epsilon_{2}\epsilon_{3}, \\\non
r&\simeq&16\epsilon_{1}\left[ 1+C\epsilon_{2}+\left(C-{\pi^{2}\over 2}+5\right)\epsilon_{1}\epsilon_{2}+\left({C^{2}\over 2}-{\pi^{2}\over 8}+1\right)\epsilon_{2}^{2}+\left({C^{2}\over 2}-{\pi^{2}\over 24}\right)\epsilon_{2}\epsilon_{3}  \right],\label{ratio}\\
\eea
where $C=\gamma_{E}+\ln 2 -2\simeq -0.7296$ and $\gamma_{E}$ is the Euler constant. 

%%%%%%%%%%%%%% BIBLIO %%%%%%%%%%%%%%%%%%%%%%%%%%

\end{document}